\documentclass[a4paper,11pt]{article}
\usepackage{jheppub} 
\usepackage[english]{babel}
\usepackage{amsmath,amssymb}

\begin{document}

\begin{flushright}
Imperial-TP-2025-CH-9\\
UUITP-25/25  
\end{flushright}

\title{The Symplectic Geometry of p-Form Gauge Fields}


\author[a]{Chris Hull}
\author[b,c]{and Maxim Zabzine}
\affiliation[a]{The Blackett Laboratory, Imperial College London, Prince Consort Road, \\
London SW7 2AZ, United Kingdom}
\affiliation[b]{Department of Physics and Astronomy, Uppsala University,\\ Box 516, SE-75120 Uppsala, Sweden}
\affiliation[c]{Centre for Geometry and Physics, Uppsala University,\\ Box 480, SE-75106 Uppsala, Sweden}
\emailAdd{c.hull@imperial.ac.uk}
\emailAdd{maxim.zabzine@physics.uu.se}
\abstract{We formulate interacting antisymmetric tensor gauge theory in a configuration space consisting of a pair of dual field strengths which has a natural symplectic structure. The field equations are formulated as the intersection of a pair of submanifolds of this infinite-dimensional symplectic configuration space, one of which is a Lagrangian submanifold while the other is either
a coisotropic or Lagrangian submanifold, depending on the topology. Chern-Simons interactions give the configuration space an interesting global structure.
We consider in detail the example of a six-dimensional  theory of a 3-form field strength coupled to Yang–Mills theory via a Chern–Simons interaction. Our approach applies
to a broad class of gauge systems.}

\maketitle

\section{Introduction}

This paper extends the geometric formulation of supersymmetric sigma models developed  in  \cite{Hull:2024gvy} to  antisymmetric tensor gauge theories. Our previous work \cite{Hull:2024gvy} sought an understanding of 
 certain geometrical constructions \cite{Bischoff:2018kzk, Alvarez:2024wki} in  generalized K\"ahler geometry in terms of the $N=(2,2)$  supersymmetric  non-linear sigma model in two dimensions \cite{Gates:1984nk}. 
 We used a \lq democratic' configuration space consisting of dual pairs of superfields and formulated the field equations and superspace constraints geometrically as the intersection of a pair of Lagrangian submanifolds of the configuration space.
 For K\" ahler target spaces, the superfields take values in Donaldson's deformation of the cotangent bundle \cite{MR1959581} while for generalized K\"ahler geometry there is an interesting generalisation of this.
  This unusual treatment 
   is closely related to the idea of symplectic realization and relates to recent approaches to the understanding of 
     generalized K\"ahler geometry. 

 Here we use a similar analysis for  simple gauge systems with fairly general dynamics, including non-linear systems such as those with Born-Infeld interactions.
 The theory of a $q-1$ form gauge field in $D$ dimensions  is formulated in terms of a $q$-form field strength  $F$ and a  $D-q$ form  dual field strength $G$. The space of fields is an infinite dimensional  symplectic manifold and we formulate the field equations and Bianchi identities as the restriction to the intersection of two submanifolds, one of which is Lagrangian while the other is  coisotropic. The coisotropic submanifold is in fact  Lagrangian if the spacetime has trivial rank-$q$ cohomology $H^q$.
In the simplest cases there is a well-defined classical action and $F$ and $G$ are globally defined forms. More generally, as in the case of Chern-Simons interactions, there are globally-defined field equations but the
action needs not be well-defined. In this case, $G$ is a locally defined differential form patched together with transition functions that are determined by the interactions, resulting in an intricate global structure.

Such democratic formulations of classical theories using a symplectic configuration space consisting of fields and dual fields seems to be applicable to many systems. For example, a similar approach to fluid dynamics is advocated in  \cite{Nekrasov}. 
We hope that our geometric democratic  approach will enhance our understanding of certain classical systems and it will be interesting to explore its implications for their quantisation.

This paper is organized as follows. In section \ref{s:general} we review the 
theory of  a $q$-form field strength $F$ in $D$ dimensions and its dual formulation in terms of   a $D-q$ form field strength $G$.
In section \ref{s:Lagrangian} we 
adopt a democratic formulation using both $F$ and $G$
and we realize the 
    equations of motion and Bianchi identities as the intersection of two  submanifolds of an infinite dimensional symplectic  configuration space.
    In section \ref{s:global} we generalise our discussion to allow Chern-Simons interactions so that the dual field $G$ is a locally defined form and we examine the resulting global structure.
 In section \ref{s:2Lagran} we enlarge the symplectic manifold and  realize the 
    equations of motion as the intersection of two Lagrangian submanifolds of the  infinite dimensional symplectic configuration space.
In section \ref{s:YM} we give a related symplectic geometry for  Yang-Mills theory.
Section \ref{s:6D} examines in detail the
non-trivial example of a 3-form field strength in 6D coupled to Yang-Mills theory  via  a Chern-Simons interaction. This model is  interesting due its global properties and illustrates the key features of our approach. 
 Section \ref{s:summary} provides a summary and some speculations on future developments.

\section{Antisymmetric tensor gauge fields}\label{s:general}

Consider a theory of $N$  $q$-form field strengths $F^a$ ($a
= 1, \ldots N$) on a space $\mathcal{M}$ which is an oriented manifold in $D$
dimensions with metric $g$ and Hodge duality operator $\ast$. The field
strengths will be taken to be closed
\begin{equation}
  d F^a = 0~,
\end{equation}
so that locally there  are $q - 1$ form potentials $A^a$ with $F^a = d A^a
.$ \ We  consider a general action
\begin{equation}
  S = \int_{\mathcal{M}} \mathcal{L} (F, \phi)~, \label{act}
\end{equation}
where $\mathcal{L} (F, \phi)$ is a $D$-form depending on the field strengths
$F^a$ and possibly on other matter fields denoted collectively by $\phi$. 
In this section and the following  section we will require that $\mathcal{L} (F, \phi)$ be a globally defined top-form, while in  section \ref{s:global} we will relax this to allow for Chern-Simons interactions.
Following \cite{Gaillard:1981rj,Cremmer:1997ct}, the
field equations for $A^a$ can be written as
\begin{equation}
  d G_a = 0~,
\end{equation}
where the $D - q$ forms $G_a$ are defined by  
\begin{equation}
  G_a = \frac{\partial \mathcal{L}}{\partial F^a}~. \label{geq}
\end{equation}
For example, if
\begin{equation}
  \mathcal{L} (F, \phi) = \frac{1}{2} N (\phi)_{a b} \, F^a \wedge \ast F^b + f
  (F, \phi)~,
\end{equation}
then
\begin{equation}
  G_a = N (\phi)_{a b} \ast F^b + \frac{\partial f}{\partial F^a}~.
\end{equation}

The action (\ref{act}) can be written in first order form as
\begin{equation}
\label{firstact}
  S = \int_{\mathcal{M}} \mathcal{L} (F, \phi) - G_a \wedge F^a~,
\end{equation}
where now $F^a$ is an unconstrained  $q$-form and $G_a = d B_a$. Then $B_a$ are lagrange
multipliers imposing $d F^a = 0$ and integrating out $B_a$ recovers the
original action (\ref{act}). The field equation for $F^a$ is (\ref{geq}) and
if this equation can be inverted to give $F$ in terms of $G$ we obtain a
dual action
\begin{equation}
  S = \int_{\mathcal{M}} \tilde{\mathcal{L}} (G, \phi)~, \label{duact}
\end{equation}
where $\tilde{\mathcal{L}} (G, \phi)$ is the Legendre transform of
$\mathcal{L} (F, \phi)$. 
The original theory with action (\ref{act}) and $F=dA$ is a theory of $A,\phi$ while the dual theory with action (\ref{duact}) and $G=dB$ is a theory of $B,\phi$.
The dual formulation will be classically equivalent to the original one provided the 
Legendre transform is well-defined and invertible, which in particular requires that $\mathcal{L} (F, \phi)$ be a convex function of $F$.
We will not restrict ourselves to such actions here, so that we will not require that there is a well-defined dual theory.
The duality of non-linear 1-form gauge theories and their relation to 
Legendre transformations was discussed in \cite{Gibbons:1995cv,Gaillard:1997rt} and the generalisation to $p$-form gauge theories was discussed in \cite{Bandos:2020hgy}.
However, our treatment is different from that in these references.

If the dual action is well-defined, it can
 itself be written in first order form as
\begin{equation}
  S = \int_{\mathcal{M}} \tilde{\mathcal{L}} (G, \phi) - G_a \wedge F^a~,
\end{equation}
where now $G_a$ is unconstrained and $F^a = d A^a$. Then $A^a$ are lagrange
multipliers imposing $d G_a = 0$ and integrating out $A^a$ recovers the dual
action (\ref{duact}). The $G$ field equation is now
\begin{equation}
  F^a = \frac{\partial \tilde{\mathcal{L}}}{\partial G_a}~.
\end{equation}

\section{A Lagrangian System}\label{s:Lagrangian}

Following \cite{Hull:2024gvy}, the field equations together with the Bianchi identities can
be reformulated as the following system of equations
\begin{equation}
  d F^a = 0~, \quad d G_a = 0~, \quad G_a = \frac{\partial \mathcal{L}}{\partial
  F^a} \label{main-system}
\end{equation}
for $q$-form fields $F^a$ and $D - q$ form fields $G_a$. Regarding the 1st
equation as a constraint implying the local existence of a potential $A^a $,
then the 2nd and 3rd equation combine to give the field equation for $A^a$
following from the action (\ref{act}). If the system is dualisable, then there
is a dual formulation
\begin{equation}
  d F^a = 0~, \quad d G_a = 0~, \quad F^a = \frac{\partial
  \tilde{\mathcal{L}}}{\partial G_a}~.
\end{equation}
Related  democratic formulations treating field strengths and their duals, and field equations and Bianchi identities, on an equal footing  have been used to give duality covariant formulations of supergravity \cite{Cremmer:1997ct}.

Our goal is to provide an  understanding of the system (\ref{main-system}) in terms of 
 symplectic geometry.  The $q$-form fields $F^a$ are sections of $(\Omega^q ({\cal M}))^N$ where
$\Omega^q ({\cal M})$ is the bundle of $q$-forms over $\mathcal{M}$, while
$G_a$ are sections of  $(\Omega^{D-q} ({\cal M}))^N$.  There is natural non-degenerate pairing 
 between these two spaces
\begin{equation}
\label{pair}
  < G, F > = \int_{\mathcal{M}} F^a \wedge G_a~,
\end{equation}
 so that they are dual to each other.

 We now
 define the symplectic structure on the space of fields to be
  $(\Omega^q ({\cal M}))^N \oplus (\Omega^{D-q} ({\cal M}))^N$
\begin{equation}\label{small-sympl}
  \omega = \int_{\mathcal{M}} \delta F^a  \wedge \delta G_a
\end{equation}
where $\delta$ is the  de Rham differential on the infinite dimensional configuration space ${\mathcal{C}} $, the space of form fields
$(F(x),G(x))$. For a proper mathematical treatment one can use the language of a variational bicomplex,  see e.g.\ \cite{Anderson1992} for further discussion. Note that for any form $\alpha$, $\delta \alpha$ has the opposite Grassmann parity to $\alpha$.

The system of equations (\ref{main-system}) define a submanifold of the configuration space  ${\mathcal{C}} $ with coordinates $(F(x),G(x))$
and we now investigate the symplectic geometry of this submanifold.
The constraint
\begin{equation}\label{minimal-Lagr}
    G_a = \frac{\partial \mathcal{L}}{\partial  F^a}
\end{equation}
 defines a Lagrangian submanifold of ${\mathcal{C}} $ and the action $\int {\cal L}$ can be understood as a generating function for this 
  Lagrangian submanifold. The restriction of the symplectic structure  $\omega$ to this submanifold vanishes  due to the graded symmetry: 
  \begin{equation}
  \omega = \int_{\mathcal{M}} \delta F^a  \wedge   \frac{\partial^2 \mathcal{L}}{\partial  F^a \partial F^b} \delta F^b =0~.
\end{equation}
  
 The constraints
 \begin{equation}\label{coi-constr}
  d F^a = 0, \quad d G_a = 0
\end{equation}
 define a coisotropic submanifold of the configuration space.
 In physics language, 
if we regard  the symplectic manifold ${\mathcal{C}} $  as a phase space, then
 (\ref{coi-constr}) are 
 first-class constraints in Dirac's terminology.
   The constraints $\int \epsilon_a dF^a$ generate the gauge transformation  $G_a \rightarrow G_a + d\epsilon_a$
   and the constraints $\int \tilde{\epsilon}^a dG_a$ generate the gauge transformation 
    $F^a \rightarrow F^a + d \tilde{\epsilon}^a$. These gauge transformations preserve the constraints (\ref{coi-constr})
    and 
    so the submanifold of ${\mathcal{C}} $ defined by (\ref{coi-constr}) is coisotropic. (See \cite{LaurentGengoux2013} for further discussion of  such submanifolds.)

Moreover, there is  a natural coisotropic reduction (which is a generalization of  sympletic reduction) in which the   constraint surface  defined by (\ref{coi-constr}) is quotiented
     by the gauge transformations (which preserve the constraint surface). The result of the coisotropic reduction 
      is again a symplectic manifold,  and in this case this is  the  cohomology space $(H^q)^N \oplus (H^{D-q})^N$ with the natural 
       symplectic structure.

  This  symplectic description treats the fields $F^a$ and $G_a$ democratically. In the coisotropic submanifold
   (\ref{coi-constr}) the fields enter on an equal footing. The Lagrangian submanifold (\ref{minimal-Lagr}) exists 
    independently  of a choice of ${\cal L}$, and if  ${\cal L}$ is convex then   the corresponding Lagrangian 
     submanifold is transverse  to  the $G$-space and $F$-space and thus it defines a map between them. 

 If the $q$-form cohomology $H^q({\mathcal{M}})=H^{D-q}({\mathcal{M}})$ is trivial, then on the  constraint surface (\ref{coi-constr}) $F$ and $G$ are exact so that $\omega$ is the integral of an exact form. For compact ${\mathcal{M}} $, or for non-compact ${\mathcal{M}} $ with suitable boundary conditions on $F,G$, this integral vanishes so that the constraint submanifold is isotropic. In this case, as it is both isotropic and coisotropic, the constraint submanifold is in fact Lagrangian.

\section{Global Structure and Chern-Simons Interactions}\label{s:global}


We now relax the condition that the Lagrangian $\mathcal{L}$ is
 a globally-defined
top form, requiring only that the field equations for $A$ and $\phi$ are globally-defined.
 Dualites for theories with Chern-Simons terms in the field strengths or in the action have been discussed in e.g.\ \cite{Cremmer:1997ct, Cremmer:1998px}; here we give a general discussion
 including the extension to non-linear theories.
 We demand  that the Lagrangian is only locally defined, so that in each patch $U_{\alpha}$ of a contractible open cover of
$\mathcal{M}$ there is a top form $\mathcal{L}_{\alpha}$, such that in
overlaps
\begin{equation}
  \mathcal{L}_{\alpha} -\mathcal{L}_{\beta} = d \rho_{\alpha \beta} \qquad
   {\rm in} \quad U_{\alpha} \cap U_{\beta}
\end{equation}
for some $D - 1$ forms $\rho_{\alpha \beta} (F, \phi)$ on $U_{\alpha} \cap
U_{\beta}$. 
 Then there are $D - q$ forms $G_{\alpha a}$ in $U_\alpha$ defined by
\begin{equation}
\label{Gis}
  G_{\alpha a} = \frac{\partial \mathcal{L}_{\alpha}}{\partial F^a}
\end{equation}
with transition functions
\begin{equation}
  G_{\alpha a} - G_{\beta a} = d \frac{\partial \rho_{\alpha \beta}}{\partial
  F^a} \qquad \rm{in} \quad U_{\alpha} \cap U_{\beta} ~.\label{trang}
\end{equation}
The $G_{\alpha a} $ are \emph{not}  globally-defined forms in general.
However, the field equations
\begin{equation}
  d G_{\alpha a} = 0 \label{dga}
\end{equation}
are well-defined as $d G_{\alpha a}$ is a globally-defined form, as $d
G_{\alpha a} = d G_{\beta a}$ in $U_{\alpha} \cap U_{\beta}$.

 An important example is that of an interaction that is linear in $F$
\begin{equation}
  \mathcal{L}_{\alpha} = \bar{\mathcal{L}} (F,\phi) + F^a \wedge \Theta_{\alpha a}~,
  \label{lcs}
\end{equation}
where $\bar{\mathcal{L}}$ is a globally-defined top form and $\Theta_{\alpha
a} (\phi)$ ($a=1,\dots, N$) are $D - q$ forms on $U_{\alpha}$ depending on the extra fields
$\phi$ but not on $F^a$, with transition functions
\begin{equation}
\Theta_{\alpha a} - \Theta_{\beta a} = d \lambda_{\alpha \beta a} \qquad
  \rm{in} \quad U_{\alpha} \cap U_{\beta}
\end{equation}
so that
\begin{equation}
  P_a = d \Theta_{\alpha a}
\end{equation}
is a globally-defined closed $D - q + 1$ form. In this case
\begin{equation}
  \rho_{\alpha \beta} = F^a \wedge \lambda_{\alpha \beta a}~.
\end{equation}
If $\mathcal{M}$ is taken to be the boundary of a $D + 1$ dimensional manifold
$\mathcal{N}$, then the action can be defined to be
\begin{equation}
  S = \int_{\mathcal{M}} \bar{\mathcal{L}} (F, \phi) + \int_{\mathcal{N}} F^a
  \wedge P_a
\end{equation}
and will give a well-defined path integral independent of the choice of
$\mathcal{N}$ provided $F^a$ and $P_a$ represent integral cohomology classes
(up to suitable constants of proportionality) so that $\int_{\mathcal{N}} F^a \wedge
P_a \in 2 \pi \hbar \mathbb{Z}$.

The $G_{\alpha a} $ are now given by 
\begin{equation}
  G_{\alpha a} = \frac{\partial \bar{\mathcal{L}}_{\alpha}}{\partial F^a} +
  \Theta_{\alpha a}~.
\end{equation}
The  $\lambda_{\alpha \beta a}$ depend only on the  fields $\phi$.
If we treat the  $\phi$  as fixed background fields, 
then $\delta \phi=0$  and we have 
  $\delta  G_{\alpha a} =\delta G_{\beta a}$ and thus the symplectic form is well-defined
(\ref{small-sympl}) but now $G$ is not  a globally defined $D-q$ form.
We can define the corrected field strengths
\begin{equation}
  \hat{G}_{\alpha a} = G_{\alpha a} - \Theta_{\alpha a}~,
\end{equation}
which are globally defined forms, as $\hat{G}_{\alpha a}=\hat{G}_{\beta a} $ in $U_{\alpha} \cap U_{\beta}$.
However,   instead of (\ref{dga}) it now satisfies
\begin{equation}
\label{modbi}
  d \hat{G}_{\alpha a} = - P_a~.
\end{equation}

The action can be written in the first order form
\begin{equation}
  S = \int_{\mathcal{M}} \left[ \bar{\mathcal{L}} (F, \phi)  - G_a \wedge F^a \right]+ \int_{\mathcal{N}} F^a
  \wedge P_a
\end{equation}
where  $F^a$ is regarded as an unconstrained  $q$-form and $G_a = d B_a$. 
This corresponds to a  local Lagrangian
\begin{equation}
  \mathcal{L}_{\alpha} = \bar{\mathcal{L}} (F,\phi) - F^a \wedge \hat{G}_{\alpha a}~,
  \label{lcsf}
\end{equation}
Eliminating $F$ gives the dual action
\begin{equation}
\label{dubiac}
  S = \int_{\mathcal{M}} \hat{\mathcal{L} }(\hat G, \phi) ~,
\end{equation}
where $\hat{\mathcal{L} }(\hat G, \phi)$ is the Legendre transform of $\bar{\mathcal{L}} (F, \phi) $.
The dual action is well-defined (when the Legendre transform is) but the Bianchi identity is now   (\ref{modbi}).

For example, for $N = 1$, a $q$-form field strength  $F $ for $q = D - 2 n + 1$
can 
be coupled to a Yang-Mills field with 2-form
field strength $\mathcal{F}$ by taking  $P =
\rm{tr} (\mathcal{F}^n)$ so that $\Theta_{2n-1}={\rm CS}_{2n-1}$ 
is the corresponding Chern-Simons $2 n -
1$ form satisfying $d {\rm CS}_{2n-1} = \rm{tr} (\mathcal{F}^n)$.
Then the local forms
 (\ref{Gis}) are
\begin{equation}
\label{Gisa}
  G_{\alpha } = \frac{\partial \bar{\mathcal{L}}_{\alpha}}{\partial F} +
  {\rm CS}_{\alpha }~,
\end{equation}
so that
\begin{equation}
  G_{\alpha } - G_{\beta } = d \lambda_{\alpha \beta } \qquad \rm{in}
  \quad U_{\alpha} \cap U_{\beta}~. \label{trans}
\end{equation}
The relation (\ref{Gisa}) implies that $G_{\alpha }$ has a non-standard transformation 
under Yang-Mills gauge transformations so as to cancel the gauge transformation of the Chern-Simons term.
The equation (\ref{Gisa}) can be rewritten as
\begin{equation}
\label{Gisaa}
  \hat G_{\alpha } = \frac{\partial \bar{\mathcal{L}}_{\alpha}}{\partial F} \end{equation}
  with 
  \begin{equation}
\label{Gisa-1}
  \hat G_{\alpha } = G_{\alpha } -
  {\rm CS}_{\alpha }~,
\end{equation}
where $ \hat G_{\alpha }$ is globally defined and  invariant under Yang-Mills gauge transformations.
We will discuss this example further in later sections.

In another example, $ \Theta_{\alpha a}$ are $q$-form gauge fields with field strengths $P_a = d \Theta_{\alpha a}$. If $P_a$ represent integral cohomology classes,  then $ \Theta_{\alpha a}$ is  a gerbe connection. As a result,  the transition relations  (\ref{trang})
of $G_{\alpha a}$ will also be those 
of a gerbe connection and   there will non-trivial relations in multiple intersections of patches.

 The discussion of 
 the previous section can be extended to this case.
 The
  system of equations (\ref{main-system}) is replaced by
\begin{equation}
  d F^a = 0~, \quad d G_{\alpha a} = 0~, \quad G_{\alpha a} = \frac{\partial \mathcal{L }_\alpha}{\partial
  F^a} \label{main-system-loc}
\end{equation}
 in which $G_a$ and $\mathcal{L }$ are no longer globally defined forms but are now \lq twisted forms' defined locally with the transition functions given above.
If, as in the examples above, $\delta  G_{\alpha a} =\delta G_{\beta a}$,  then the symplectic form is well-defined on the configuration space and the system of equations (\ref{main-system-loc}) defines the intersection of a Lagrangian submanifold of the configuration space with a coisotropic one.
Here we are treating the extra fields $\phi$ as background fields on which $\delta$ does not act. In the following sections we will extend our formalism to treat the case in which all fields are dynamical and $\delta$ acts on all of them.
For the dual action (\ref{dubiac}), the system of equations would be
\begin{equation}
  d F^a = 0~, \quad d \hat G_{\alpha a} = -P_a~, \quad F^{ a} = \frac{\partial \hat {\mathcal{L }}
  }{\partial
  \hat G_a}~, \label{main-system-loc-1}
\end{equation}
which has a configuration space of globally  defined forms $F^a, \hat G_a$ but with a modified Bianchi identity for $\hat G_a$.

\section{An extended configuration space}\label{s:2Lagran}

 
We now return to the system described in section \ref{s:general} with   action 
\begin{equation}
  S = \int_{\mathcal{M}} \mathcal{L} (F, \phi)~, 
  \label{actor}
\end{equation}
 where  the Lagrangian is globally defined and  $F^a = d A^a$ with $A_a$    connection $q-1$ forms.
 The  equations of motion are
 \begin{equation}\label{eq-motion-4}
 d \Bigg (\frac{\partial {\cal L}}{\partial F^a} \Bigg ) =0 
 \end{equation}
 and the Bianchi identity is $dF^a=0$. 
 
 We now enlarge the configuration space by introducing further fields so that we can interpret our system as an intersection of two Lagrangian submanifolds.  In addition to the $q$-form fields $F^a$ and $G_a$  ($a=1,\dots, N)$, we have  $(q-1)$ form connections $A^a$
 and $(D-q+1)$ forms $C_a$.
 This enlarged configuration space is equipped with the symplectic structure
\begin{equation}\label{Big-symplec}
    \omega= \int_{\cal M} \Big ( \delta A^a \wedge \delta C_a + \delta F^a \wedge \delta G_a \Big )~.
\end{equation}
For the moment we take $F^a$,  $G_a$ and $C_a$ to be globally defined forms while
the $A^a$ are interpreted as
    $(q-1)$ form connections with transition functions
    \begin{equation}
      A^a_\alpha - A^a_\beta = d \gamma_{\alpha\beta}~,\qquad \rm{in}
  \quad U_{\alpha} \cap U_{\beta}~,
    \end{equation}
    As the  $\gamma_{\alpha\beta}$ are independent of all fields it follows that $\delta \gamma_{\alpha\beta}=0$ so that
      $\delta A^a_\alpha = \delta A^a_\beta$. Then $\delta A^a$ is globally defined and so the  symplectic structure (\ref{Big-symplec}) is well-defined.

We now define two Lagrangian submanifolds of this enlarged configuration space.
 The first Lagrangian submanifold is defined by the conditions
\begin{equation}
    F^a = d A^a~,~~~~~C_a = (-1)^{q-1} d G_a~,
\end{equation}
and is generated by  the action
\begin{equation}
   \int_{\mathcal{M}} 
  G_a \wedge dA^a
  ~, 
  \label{actory}
\end{equation}
 and the  second  Lagrangian submanifold is defined by
 \begin{equation}
     G_a = \frac{\partial {\cal L}}{\partial F^a}~,~~~~C_a=0~,
 \end{equation}
and is  generated by  the action (\ref{actor}).
  The intersection of these two Lagrangian submanifolds corresponds 
   to the equations (\ref{eq-motion-4}).

 The properties of the symplectic structure (\ref{Big-symplec}) depend on  $D$ and $q$. For example, for the case $N=1$ and $D=2q -1$ we have $\deg A = \deg G$ and $\deg C = \deg F$ and  the symplectic structure (\ref{Big-symplec}) can be rewritten as
   follows
   \begin{equation}
       \omega= \frac{1}{2} \int_{\cal M} \Big ( \delta (A+i F) \wedge \delta (C-iG) + \delta (A - i F) \wedge \delta (C+iG)
        \Big )
   \end{equation}
  and so it corresponds to a real part of holomorphic symplectic structure. 
  Thus in the case $D=2q-1$ the space of fields is equipped with a holomorphic symplectic structure.  
  
  The above can be generalised to the case in which $\mathcal{L}$ and  $G_a$ are only locally defined. The discussion then follows closely that of the previous section.

\section{$D$-dimensional Yang-Mills theory}\label{s:YM}

As a preparation for the following section in which we discuss $q-1$ form gauge fields interacting with Yang-Mills theory, we now
 consider  Yang-Mills theory in $D$ dimensions. The action is  
\begin{equation}
   S_{\rm YM} (a) = \frac{1}{2}\int_{\cal M} {\rm Tr} ({\cal F}(a) \wedge \star {\cal F}(a) )~,   
\end{equation}
 where ${\cal F}(a) = d a + a^2$ is the field strength and $a$ is connection taking values in a  quadratic Lie algebra $\mathfrak{g}$ with invariant pairing given by ${\rm Tr}$ and corresponding gauge group $\mathbf{G}$. The above action can be rewritten in the first order
  formalism as follows
  \begin{equation}
       S_{\rm YM} (a,K)= \int_{\cal M} {\rm Tr} \Big ( K (da + a^2) - \frac{1}{2} K \wedge \star K \Big )~,
  \end{equation}
   where $K$ is an adjoint-valued $D-2$ form. The equations of motion are 
   \begin{equation}\label{EM-YM-first}
     \star K = da + a^2~,~~~~ d_a K = dK + [a,K]=0~,  
   \end{equation}
 where we assume the metric has Euclidean signature for concreteness. Substituting $\star K = da + a^2$ in the action (\ref{EM-YM-first}) recovers the original action.

 Let us reformulate the above equations of motion as the intersection of two Lagrangian submanifolds in an infinite dimensional symplectic space. The space of fields consists of the connection $a$, an adjoint valued $D-1$ form  $b \in \Omega^{D-1}({\cal M}, \mathfrak{g})$,
an adjoint valued 2-form  ${\cal F} \in \Omega^{2}({\cal M}, \mathfrak{g})$  and an adjoint valued $D-2$ form $K \in \Omega^{D-2}({\cal M}, \mathfrak{g})$. Note that here ${\cal F}$ is an arbitrary 2-form and we don't require that it satisfies a Bianchi identity.
These fields transform under the gauge transformations 
\begin{eqnarray}
 &&   a~\rightarrow~g^{-1}a g+ g^{-1} dg~, \label{gauge-transformations-1}
 \nonumber
  \\
  &&  b~\rightarrow~g^{-1} b g~, \label{gauge-transformations-2}
  \nonumber\\
   &&  {\cal F}~\rightarrow~g^{-1}{\cal F} g~, \label{gauge-transformations-3}
   \nonumber \\
     && K~\rightarrow~g^{-1} K g~ \label{gauge-transformations-4} 
\end{eqnarray}
with $g\in\mathbf{G}$.
This space of  fields
  is equipped with the   symplectic structure 
 \begin{equation}
     \omega = \int\limits_{\cal M}~{\rm Tr}  \Big ( \delta a \wedge \delta b + \delta {\cal F} \wedge \delta K \Big )  ~,
 \end{equation}
 which is invariant under the  gauge transformations (\ref{gauge-transformations-4}).

   The first Lagrangian submanifold is defined by the conditions
   \begin{equation}
       {\cal F} = da + a^2~,~~~~b= - dK- [a, K] = - d_{a} K
   \end{equation}
   and the second Lagrangian submanifold is defined by the conditions
   \begin{equation}
     {\cal F} = \star K~,~~~~b=0~.   
   \end{equation}
    The intersection of these two Lagrangian submanifolds gives us the equations of motion for the Yang-Mills system (\ref{EM-YM-first}). 
  The first Lagrangian submanifold is generated by
  \begin{equation}
   \int\limits_{\cal M}~{\rm Tr}  \Big ( [ da + a^2]\wedge K
   \Big )  
   \end{equation}
while the second is generated by
\begin{equation}
  \frac 1 2 \int\limits_{\cal M}~{\rm Tr}  \Big (K\wedge \ast K
   \Big )  \, .
   \end{equation}

  \section{A 6-dimensional system}\label{s:6D}

  We now  discuss a 6-dimensional system which provides a very good illustration of our general ideas and
   shows the intricate relation between infinite dimensional symplectic geometry and global issues.  
  We choose 6 dimensions 
   for the sake of concreteness and there are straightforward generalisations to other systems in other dimensions.   

  Consider the theory of a closed 3-form field strength $F=dA$, where $A$ is a 2-form potential,  coupled 
  to Yang-Mills theory via a Chern-Simons interaction 
  \begin{equation}\label{6D-full-action}
        S = \int_{\mathcal{M}} \Big ( \bar{ \mathcal{L} }(F, \phi) + F \wedge {\rm CS}_3(a) \Big ) + S_{\rm YM} (a)~,
  \end{equation}
  with $ \bar{ \mathcal{L} }(F, \phi) $ a globally-defined gauge-invariant top-form.
Here the Chern-Simons term is   \begin{equation}
    {\rm CS}_3(a) = {\rm Tr} \Big ( a d a + \frac{2}{3} a^3 \Big ) 
 \end{equation}
  with the following gauge transformations
 \begin{equation}\label{transf-CS}
     {\rm CS}_3 \Big (g^{-1} a g+ g^{-1} dg \Big ) ={\rm CS}_3 (a) + d {\rm Tr} (g^{-1}dg~ a) - \frac{1}{3} {\rm Tr} \Big (
     (g^{-1} dg)^3 \Big )~
 \end{equation}
 under $a~\rightarrow~g^{-1}a g+ g^{-1} dg$.
  Then $e^{\frac{i}{4\pi}S}$  is gauge invariant provided that 
  $F \in H^3 ({\cal M}, \mathbb{Z})$. 
   The equations of motion for this system are given by the following expressions
  \begin{equation}\label{6D-EqM}
 d \Bigg( \frac{\partial \bar{{\cal L}}}{\partial F} \Bigg) + {\rm Tr} \Big ({\cal F}(a) \wedge {\cal F}(a)\Big )=0~,~~~~
  d_a \star {\cal F}(a) - 2 F \wedge {\cal F}(a)=0
 \end{equation}
 together with the Bianchi identities $dF=0$ and $d_{a} {\cal F}(a)=0$. Recall that $A$ and $F$ are invariant under the gauge group. 
 If the Yang-Mills field $a$ is treated as a background field, then this system can be treated as explained in 
 section \ref{s:global}.
 Our goal here is to treat the Yang-Mills fields as  dynamical fields, treated in the way explained  in section \ref{s:YM}, and to
 interpret the equations of motion as the intersection of two Lagrangian 
   submanifolds of the infinite dimensional symplectic manifold which is the configuration space for the combined system of the $2$-form gauge field and  the Yang-Mills fields. 

 We start by combining the discussions from sections \ref{s:2Lagran} and \ref{s:YM}. In addition to the 3-form $F$ and the
  $2$-form potential $A$, we introduce  corresponding dual fields which are also gauge singlets: a 3-form $G$ and 4-form $C$. On the Yang-Mills side, 
   in addition to the connection $a$,  we introduce  additional forms in the  adjoint representation: a 5-form $b$, a 2-form ${\cal F}$
    and a $4$-form $K$. We  define the following symplectic form on the space of fields
   \begin{equation}
   \label{SymSt}
     \omega = \int\limits_{\cal M}~\Big [\delta A \wedge \delta C + \delta F \wedge \delta G+ {\rm Tr}  \Big ( \delta a \wedge \delta b + \delta {\cal F} \wedge \delta K \Big )  \Big ]~.
 \end{equation}

  In this new symplectic space the equations of motion (\ref{6D-EqM}) correspond to the intersection of two Lagrangian submanifolds. 
The first Lagrangian manifold is defined by the equations
\begin{equation}
\label{firLag}
    F= dA~,~~~~C=(-1)^{q-1}dG~,~~~~b= - d_a K~,~~~~{\cal F} = d a + a^2~
\end{equation}
and is generated by
\begin{equation}
  \int_{\mathcal{M}} 
    G_a \wedge dA^a +{\rm Tr} ( [ da + a^2]\wedge K)
  ~.
  \label{aactory}
\end{equation}
 The second Lagrangian submanifold is defined by the equations
\begin{equation}
\label{secLag}
  G = \frac{\partial \bar{{\cal L}}}{\partial F} + {\rm CS}_3 (a)~,~~~~b= 2 F (da+a^2) - dF a~,~~~~{\cal F}= \star K~,
\end{equation}
together with $C=a=A=F={\cal F}=0$
and is generated by
 \begin{equation}
   \int\limits_{\cal M}~{\cal L}+ 
 \frac 1 2  {\rm Tr}  \Big (K\wedge \ast K
   \Big ) 
   \end{equation}

 The equation $G - {\rm CS}_3 (a)= \frac{\partial \bar{{\cal L}}}{\partial F} $ in (\ref{secLag}) is only invariant under Yang-Mills gauge transformations if 
 $G$ transforms in such a way that its variation cancels the  
  transformation (\ref{transf-CS}) 
 of the Chern-Simons term. This requires that $G$ transforms as  
 \begin{equation}
G~\rightarrow~G + d {\rm Tr} (g^{-1} dg~ a) - \frac{1}{3} {\rm Tr} \Big (
     (g^{-1} dg)^3 \Big )~,\label{new-tr-G}
\end{equation}
However, this transformation combined with the gauge transformations (\ref{gauge-transformations-4}) does not preserve the symplectic form (\ref{SymSt}), but a symplectomorphism is achieved by further modifying the gauge transformation of $b$ to become $b~\rightarrow~ g^{-1}b g - (g^{-1} dg) ~ dF$.
As a result, the symplectic form and the Lagrangian submanifolds are preserved under the gauge transformations
\begin{eqnarray}
&& G~\rightarrow~G + d {\rm Tr} (g^{-1} dg~ a) - \frac{1}{3} {\rm Tr} \Big (
     (g^{-1} dg)^3 \Big )~,\nonumber \\
 &&   a~\rightarrow~g^{-1}a g+ g^{-1} dg~, 
  \nonumber
  \\
  &&  b~\rightarrow~g^{-1} b g- (g^{-1} dg) ~ dF,   \nonumber\\
   &&  {\cal F}~\rightarrow~g^{-1}{\cal F} g~,
   \nonumber \\
     && K~\rightarrow~g^{-1} K g~ \label{gauge-transformations-5} 
\end{eqnarray}
with the fields $A$, $C$, $F$    invariant under the gauge group.

In each contractible chart $U_\alpha$ of $\cal{M}$ we have differential forms $A,C,F,G,a,b,{\cal F} ,K$ and we glue these together on intersections $U_\alpha\cap U_\beta$ via gauge transformations (\ref{gauge-transformations-5}) to define a symplectic configuration space with a globally defined symplectic structure. 
Then with these transition functions the conditions (\ref{firLag})  and (\ref{secLag}) define two well-defined Lagrangian submanifolds.
The above deformation of the transition functions for $G,b$ can be viewed as an  
 infinite dimensional analogue of  Donaldson's twist of the cotangent bundle of 
  a K\"ahler manifold which is  necessary to define an appropriate Lagrangian submanifold for the sigma-model configuration space, see \cite{Bischoff:2018kzk,Hull:2024gvy}.

\section{Summary}\label{s:summary}

This note builds on our previous work \cite{Hull:2024gvy}, extending the approach we used for sigma models to  gauge systems. 
We use a democratic configuration space of  dual pairs of fields 
and interpret the equations of motion as intersections of Lagrangian or   coisotropic submanifolds of the 
infinite-dimensional symplectic configuration space.
We believe that our ideas can be naturally extended to a broader class of 
 models.
 
The intersection of two  Lagrangian submanifolds might be thought of as 
being generated by two different \lq Hamiltonians' or as
the classical field equations resulting from simultaneously extremising two different actions. It will be interesting to explore this viewpoint  and its generalisations further.

Our approach has been classical, but we hope that these ideas might lead to an  improvement our understanding of  
quantisation and in particular  of duality in quantum theories.
 In the present setting,  our formulation is democratic and geometric and could be a good starting point for a geometric quantisation. 
 For example,
   we may try to quantise some 
  coisotropic/Lagrangian submanifolds in some formal geometrical quantisation framework. 
One approach could be to formulate the theory on a  $D+1$ dimensional space whose boundary is the $D$-dimensional spacetime, with the coisotropic condition  coming from a  $D+1$ dimensional topological theory 
while the other Lagrangian submanifold encodes a boundary condition for this theory. 
    We leave these and related questions for the future. 


 \bigskip\bigskip
\noindent{\bf\Large Acknowledgement}:
\bigskip

\noindent We are grateful to Nikita Nekrasov for illuminating discussions. 
 The research of CH is supported by   the STFC Consolidated Grant ST/T000791/1.  The research of MZ is  supported by the Swedish
Research Council 
excellence center grant ``Geometry and Physics'' 2022-06593.
  This work is also supported by the Swedish
Research Council under grant no. 2021-06594 while both authors were in residence at the Institut Mittag-Leffler in Djursholm, Sweden during January-April
of 2025.

\bibliographystyle{utphys}
\bibliography{references}{}

\end{document}